\documentclass[aps,preprint,superscriptaddress,journal=prl]{revtex4-2}
\usepackage{amsmath,amsthm,amssymb,amsfonts}
\usepackage{siunitx}
\usepackage{graphicx} 
\usepackage{color}
\usepackage{hyperref}
\usepackage{diagbox}
\usepackage{siunitx}
\usepackage{textcomp}
\usepackage{gensymb}
\hypersetup{colorlinks=true,linkcolor=blue,anchorcolor=blue,citecolor=blue}

\begin{document}

\title{\textbf{Giant Reversible Piezoelectricity from Symmetry-Governed Stochastic Dipole Hopping}}

\author{Denan Li}
\affiliation{Fudan University, Shanghai 200433, China}
\affiliation{Department of Physics, School of Science, Westlake University, Hangzhou, Zhejiang 310030, China}

\author{Haofei Ni}
\affiliation{Institute for Science and Applications of Molecular Ferroelectrics, Key Laboratory of the Ministry of Education for Advanced Catalysis Materials, Zhejiang Normal University, Jinhua, Zhejiang 321004, China}

\author{Yi Zhang}
\affiliation{Institute for Science and Applications of Molecular Ferroelectrics, Key Laboratory of the Ministry of Education for Advanced Catalysis Materials, Zhejiang Normal University, Jinhua, Zhejiang 321004, China}

\author{Shi Liu}
\email{liushi@westlake.edu.cn}
\affiliation{Department of Physics, School of Science, Westlake University, Hangzhou, Zhejiang 310030, China}
\affiliation{Institute of Natural Sciences, Westlake Institute for Advanced Study, Hangzhou, Zhejiang 310024, China}

\begin{abstract}
Organic--inorganic hybrid perovskites with giant piezoelectric responses, exemplified by TMCM-CdCl$_3$, represent a promising platform for flexible and environmentally friendly electromechanical materials. However, the microscopic origin of such exceptional performance in this weakly polar system has remained elusive.
Here, using deep-learning-assisted large-scale molecular dynamics simulations, we resolve this paradox by reproducing the experimentally measured piezoelectric coefficient $d_{33} \approx 220$~pC/N, and demonstrating that the giant response arises from the collective contribution of multiple intrinsic components, particularly the shear component $d_{15}$. This effect does not stem from conventional polarization rotation or phase switching, but instead originates from stochastic 120$^\circ$ in-plane rotational hopping of a small fraction of organic cations. This discrete hopping mechanism is governed by the local C$_3$-symmetric halogen-bonding network between the host framework and the guest cation.
The Arrhenius-type temperature dependence of $d_{15}$ further confirms the role of thermally activated dipole hopping. This work provides a clear pathway to enhance piezoelectric performance of hybrid materials through rational engineering of host--guest interactions.
\end{abstract}

\maketitle
\clearpage
Piezoelectric materials, which can convert mechanical energy into electrical energy and vice versa, are foundational to a wide range of modern technologies, from ultrasound to advanced sensors~\cite{Li25peadn4926}. For decades, this technological landscape has been dominated by inorganic ferroelectrics such as barium titanate (BTO) and lead zirconate titanate (PZT)~\cite{Acosta17p041305,Marion21p961,Jaffe54p809,Zhang15p1,Li23p87}. While these materials offer robust performance, they also present notable drawbacks: they are heavy, brittle, require high-temperature processing, and often contain toxic elements, making them unsuitable for emerging applications that demand flexible, lightweight, and biocompatible piezoelectrics.
This has driven a search for molecular alternatives, which promise mechanical flexibility and environmentally friendly solution-processability, but have historically been limited by a fundamental trade-off: molecular ferroelectrics tend to exhibit low spontaneous polarization, leading to weak piezoelectric response~\cite{Lovinger83p1115,Liu20p1902468,Sun16p11854,Ye18p151,Liao19p6432}. 

The discovery of the hybrid organic-inorganic ferroelectric perovskite trimethylchloromethyl ammonium trichlorocadmium(II) [Me$_3$NCH$_2$ClCdCl$_3$, (TMCM-CdCl$_3$)], composed of organic TMCM$^+$ cations and inorganic CdCl$_3^-$ frameworks, marks an important advancement~\cite{You17p306}. Exhibiting a giant piezoelectric coefficient, it challenges the long-standing trade-off typically associated with molecular ferroelectrics. This breakthrough opens up new opportunities for developing flexible, low-temperature-processable piezoelectric materials.
However, it also presents a fundamental paradox: despite having a spontaneous polarization of only 4--5~$\mu$C/cm$^2$, an order of magnitude smaller than that of BTO ($\approx$25$\mu$C/cm$^2$), TMCM-CdCl$_3$ achieves a piezoelectric strain coefficient of $d_{33}\approx220$~pC/N, comparable to BTO. The striking anomaly of a weakly polar material exhibiting a giant piezoelectric response suggests the presence of an unconventional physical mechanism.

Two primary theories have been proposed to resolve the paradox of giant piezoelectricity in weakly polar TMCM-based perovskites. The original study suggested a ferroelastic polarization rotation mechanism~\cite{You17p306}. Crystallographic and vector piezoresponse force microscopy analyses revealed that monoclinic TMCM-CdCl$_3$ hosts twelve energetically equivalent polar states. It was argued that applied stress could trigger ferroelastic switching between these states, causing a large-angle reorientation of the polarization vector and, consequently, a giant piezoelectric response.
A key limitation of this theory is that reversible ferroelastic switching, essential for reversible piezoelectricity, often requires alternating compressive and tensile stress, a condition not typically met during standard unipolar piezoelectric measurements.
A later theoretical study proposed a phase-switching mechanism~\cite{Ghosh20p207601}. First-principles calculations identified a metastable phase featuring TMCM$^+$ cations in head-to-head and tail-to-tail orientations. The piezoelectric response was attributed to a stress-induced transition between this phase and the ground-state monoclinic structure.
This theory was based on static, zero-Kelvin unit-cell calculations that neglect thermal fluctuations and collective dipole dynamics, factors likely crucial to understanding the experimental behavior at finite temperatures and in mesoscale systems.

Here, we resolve this long-standing puzzle using large-scale molecular dynamics (MD) simulations of TMCM-CdCl$_3$, enabled by a high-fidelity machine-learning force field trained exclusively on first-principles data. By computing the full piezoelectric tensor in the single-domain limit at 300~K, we show that the experimentally observed giant response is dominated not by the intrinsic longitudinal coefficient $d_{33}$, but by large shear components, particularly $d_{15} = 163$~pC/N. Notably, applying a proper tensor transformation from the crystal to the laboratory frame yields an effective $d^{\rm eff}_{33}$ of 212 pC/N at 300 K, closely matching the experimental value of 220 pC/N.
We uncover a novel piezoelectric mechanism distinct from both ferroelastic and phase switching: a stress-driven, thermally-activated stochastic 120$^\circ$ rotational hopping of a small fraction of TMCM$^+$ cations. This hopping is governed by the local C$_3$ symmetry imposed by the halogen bonds between organic cations and the inorganic framework. 
As long as the fraction of hopped cations remains below a critical threshold, they can return to their original states once the stress is removed, exhibiting a shape-memory-like effect. This collective, reversible dipole hopping underlies the observed reversible piezoelectricity, further supported by the strong temperature dependence of the response. Finally, we distinguish this useful, reversible behavior from the irreversible, large-strain ferroelastic switching that defines the material’s fatigue limit. Our results provide a comprehensive picture of piezoelectricity in TMCM-CdCl$_3$ and similar hybrid perovskites, offering key insights for further improving electromechanical responses in these materials.

\begin{figure*}
\centering
\includegraphics[width=1\textwidth]{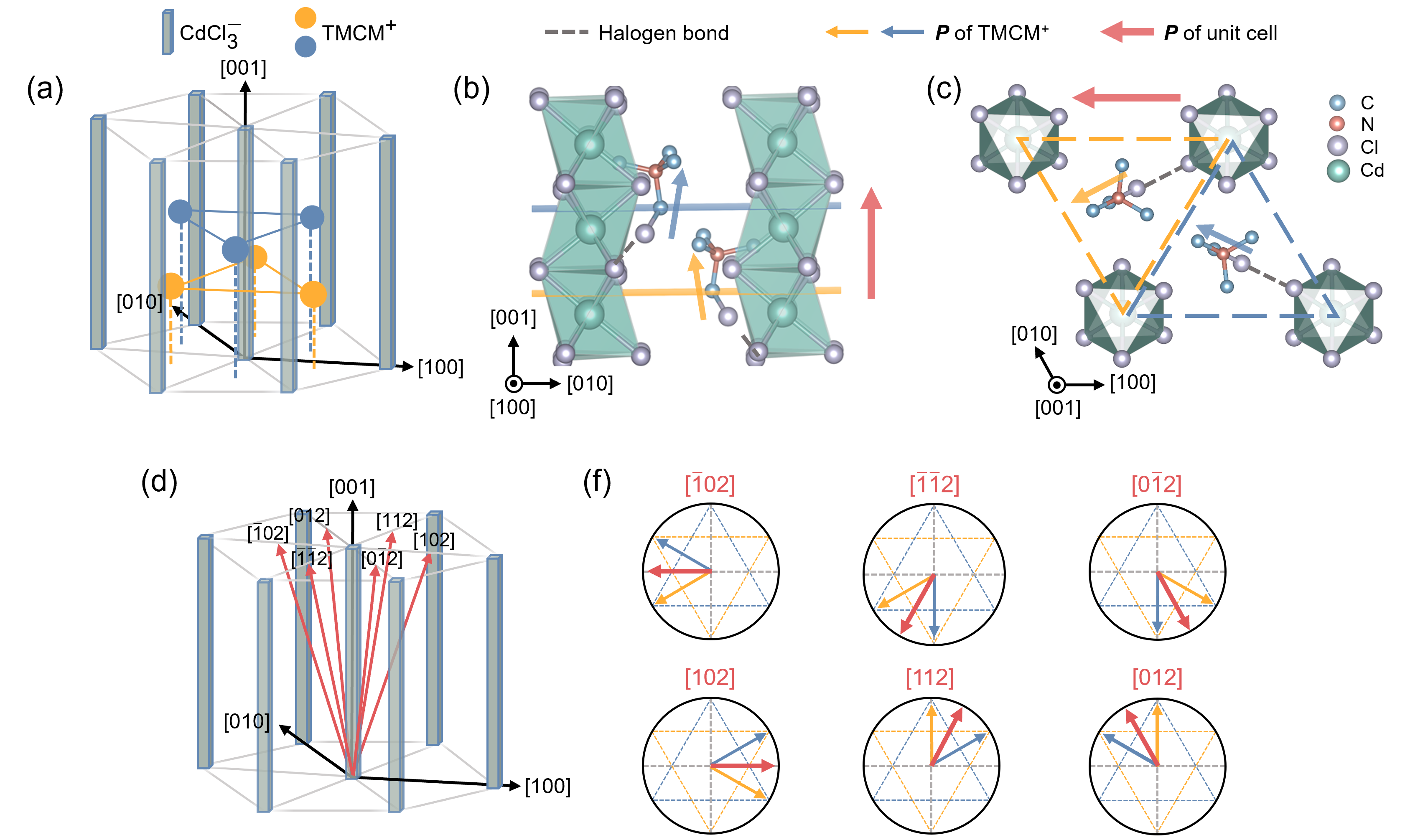}
\caption{
\textbf{Crystal structure and polarization varients of TMCM-CdCl$_3$.} 
(a) Schematic illustration of the host–guest framework in TMCM-CdCl$_3$. One-dimensional inorganic CdCl$_3^-$ chains form a triangular lattice, with organic TMCM$^+$ cations occupying the interstitial sites to create A–B stacked triangular sub-lattices. The crystallographic axes and directions shown are defined with respect to the paraelectric hexagonal lattice. (b) Side view and (c) top view of the crystal structure in the ferroelectric phase. The dashed triangle indicates the local triple-well potential energy surface associated with the in-plane rotation of a TMCM$^+$ cation, which exhibits C$_3$ symmetry. In this phase, molecules are anchored by highly directional CH$_3$--Cl$\cdots$Cl halogen bonds (gray dashed lines). Due to the helical arrangement of Cl atoms along the CdCl$^-$ chain, the molecular dipoles in adjacent layers are not fully aligned.   
(d) Six equivalent macroscopic polarization directions in the ferroelectric phase, each with different in-plane components. 
(f) Schematics showing the collective in-plane orientations of molecular dipoles corresponding to the six polar variants in (d). Yellow and blue triangles represent the two dipole layers within the unit cell as shown in (b) and (c), and red arrows denote the net in-plane polarization.
}
\label{fig:fig1}
\end{figure*}

Previous studies have shown that TMCM-CdCl$_3$ crystallizes in the monoclinic space group $Cc$ in the ferroelectric phase and the hexagonal space group $P6_3/ mmc$ in the paraelectric phase. 
Despite several DFT-based investigations, the connection between the collective orientations of molecular dipoles associated with TMCM$^+$ cations and the macroscopic polarization remains controversial in literature~\cite{Ghosh20p207601,Lin25p1,Kasper23p8885}. We introduce an intuitive host-guest framework to clarify this relationship. As depicted in Fig.~\ref{fig:fig1}a, one-dimensional CdCl$_3^-$ chains, the ``host," self-assemble into a triangular lattice, forming triangular channels along the [001] direction.
The functional ``guests" are organic TMCM$^+$ cations, which reside within these channels and form their own A-B stacked triangular sub-lattices. 
In the paraelectric phase, the molecular cations are fully disordered, yielding a primitive hexagonal unit cell. 
In contrast, the ferroelectric phase features alignment of the out-of-plane components of all molecular dipoles (Fig.~\ref{fig:fig1}b).
However, a previously underappreciated detail is that, while dipoles within each layer are aligned, adjacent layers of dipoles are not fully co-aligned, as seen in the side view (Fig.~\ref{fig:fig1}b) and top-down view (Fig.~\ref{fig:fig1}c).
We find that the orientation of each TMCM$^+$ cation is anchored by highly directional CH$_3$--Cl$\cdots$Cl halogen bonds (see dashed lines in Fig.~\ref{fig:fig1}b-c), which act as a ``molecular lock" connecting the guest cation to the surrounding inorganic chains. This interaction restricts the cation to one of three energetically equivalent orientations, with its intrinsic (in-plane) dipole pointing toward an edge of the triangular channel. 
The helical arrangement of Cl atoms along the CdCl$_3^-$ chains therefore prevents full dipole alignment across layers. 

Since the unit cell of the ferroelectric phase contains two layers of molecular dipoles, the in-plane component of the total polarization can be intuitively understood using the schematic in Fig.~\ref{fig:fig1}f, which represents each dipole layer as a triangle, with the two triangles rotated 180$\degree$ relative to each other to reflect the stacking arrangement. 
The six macroscopic polar variants along the [102] and other crystallographically equivalent directions, illustrated in Fig.~\ref{fig:fig1}d, correspond to distinct combinations of allowed relative orientations between the two layers of TMCM$^+$ cations within the unit cell. As discussed in detail below, the restricted molecular dipole orientations, governed by the C$_3$ symmetry, play a key role in the piezoelectric response.

Theoretical prediction of finite-temperature piezoelectric coefficients in hybrid materials such as TMCM-CdCl$_3$ is challenging, as conventional approaches such as density functional perturbation theory often yield negligible values~\cite{Ghosh20p207601,Sodahl23p729}. This limitation likely arises from the neglect of thermal fluctuations and the associated order--disorder behavior of molecular dipoles. To overcome this challenge, we developed a deep neural network-based interatomic potential (deep potential, or DP~\cite{Zhang18p143001,Zhang18p4441}), which enables large-scale MD simulations of TMCM-CdCl$_3$ at finite temperatures. It can also directly predict total polarization from atomic positions.
Our model demonstrates high fidelity across a wide range of validation tests. It achieves an excellent fit to the reference DFT database, with low mean absolute errors (MAEs) for energy (0.214~meV/atom), force (0.017~eV/\AA), virial (1.609~meV/atom), and polarization (0.06 $\mathrm{\mu C/cm^2}$). The model accurately reproduces static properties, including the equation of state and elastic moduli, and captures key dynamic and finite-temperature phenomena, such as the energy barrier for in-plane TMCM$^+$ cation rotation and the temperature-driven ferroelectric-to-paraelectric phase transition (see Supplementary Material for details,~\cite{Chen10p445501,Li16p503,Lin24pe1687,Schlipf15p36,Perdew96p3865,Thompson22p108171,Grimme10p154104,Zeng23p054801,Zhang20p107206,Tribello14p604,Laio02p12562,Wu21p024108,Wu23p144102}). 
The piezoelectric tensor element is calculated as $d_{ij} = \Delta P_i / \sigma_j$, reflecting the polarization change along direction $i$ ($\Delta P_i$) due to an applied stress with component $j$ ($\sigma_j$ in Voigt notation). 

\begin{figure*}
\centering
\includegraphics[width=1\textwidth]{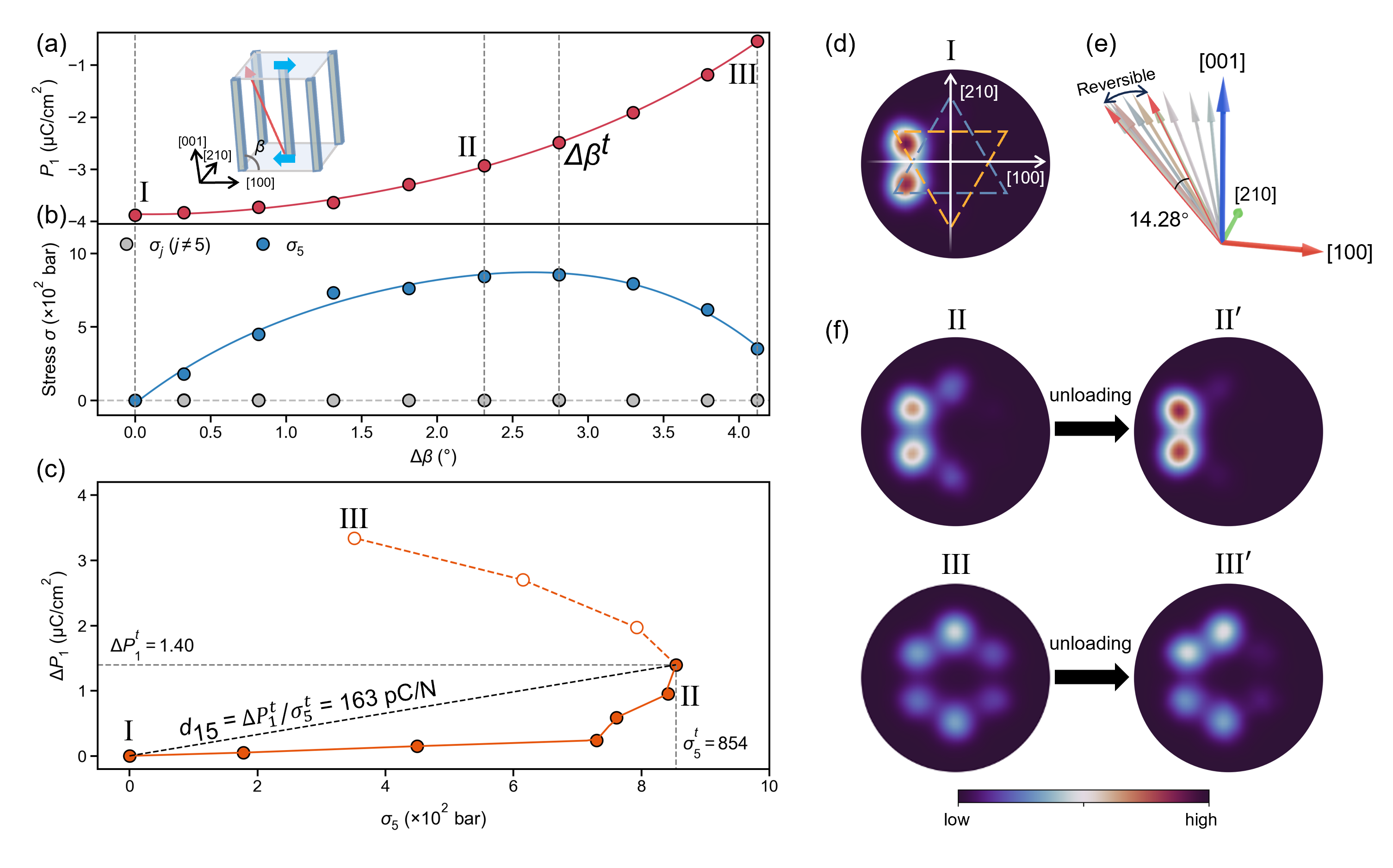}
\caption{
\textbf{Shear stress-induced reversible giant piezoelectric response via molecular dipole hopping.} 
(a) Evolution of the polarization component along [100], $P_1$, and (b) shear stress $\sigma_5$ as a function of shear strain $\Delta \beta$. The inset in (a) illustrates the initial state I, characterized by a monoclinic unit cell with in-plane polarization aligned along [$\bar{1}$00]. The shear strain is imposed by varying the lattice angle $\beta$ while allowing all other lattice parameters to relax.
MD simulations ensure that for a fixed $\Delta \beta$, all other stress components ($\sigma_j$, $j \ne 5$) remain negligible, as confirmed in (b). The shear stress $\sigma_5$ reaches a maximum at a critical strain $\Delta \beta^t$. (c) Polarization change $\Delta P_1$ as a function of $\sigma_5$, derived from (a) and (b). (d) In-plane dipole orientation density plot for TMCM$^+$ cations in the initial state I, following the schematic in Fig.~1. (e) 
Reversible region for the rotation of the macroscopic polarization. $\Delta \beta_t$ corresponds to a small rotational angle of  $14.28\degree$.   
(f) Dipole orientation density plots for strained states II and III, and their corresponding relaxed configurations after unloading (II$'$ and III$'$). The transition between states I and II is reversible, while the transition from I to III is irreversible as III$'$ is different from I.
}
\label{fig:fig3}
\end{figure*}

We now explain the calculation of $d_{15}$ (with indices defined relative to the monoclinic unit cell) in detail.  
A shear deformation is imposed by varying the lattice angle $\beta$ by an amount $\Delta \beta$ (Fig.~\ref{fig:fig3}a inset), while allowing all other lattice parameters to fully relax during finite-temperature MD simulations with $\beta$ held fixed, using a 10,800-atom supercell. At equilibrium, only the shear stress component $\sigma_5$ is nonzero, establishing a one-to-one relationship among $\Delta \beta$, $\sigma_5$, and $P_1$.
The evolution of $P_1$ and $\sigma_5$ as a function of $\Delta \beta$ at 300~K is shown in Fig.~\ref{fig:fig3}, along with dipole density plots that illustrate the distributions of in-plane molecular dipole orientations within the supercell. The starting polar state I has a net in-plane polarization along the [$\overline{1}$00] direction ($P_1 \approx -4~\mu$C/cm$^2$).
As shear strain is applied, the shear stress $\sigma_5$ increases in magnitude, while all other stress components ($\sigma_j$, $j \ne 5$) remain negligible (Fig.~\ref{fig:fig3}b). This confirms that in MD simulations, the system reaches mechanical equilibrium under the constraint of fixed $\Delta \beta$, with $\sigma_5$ representing the balanced internal stress required to maintain the imposed deformation.
The microstructural evolution under shear is revealed by the dipole density plots (Fig.~\ref{fig:fig3}f). In the strained state II, MD simulations show that dipoles do not rotate uniformly. Instead, a subset of molecular dipoles undergo 120$^\circ$ hops into adjacent symmetry-equivalent potential wells. These locally reoriented dipoles act as ``orientational defects'' embedded within the original domain~\cite {Kasper23p8885}.

Notably, the shear stress $\sigma_5$ reaches its maximum value, $\sigma_5^t$, at a critical shear deformation $\Delta \beta^t\approx2.8\degree$, marking the threshold for piezoelectric reversibility. State II lies just below this threshold. When the stress is removed, the system relaxes to state II$'$ (see Fig.~\ref{fig:fig3}f), which is nearly identical to the initial state I. This indicates that dipoles return to their original orientations via reverse 120$^\circ$ hopping, resulting in a fully reversible piezoelectric response.
In contrast, when the applied shear strain exceeds the threshold ($\Delta \beta > \Delta \beta^t$), the system transitions to a fundamentally different configuration, represented by state III, characterized by a more isotropic distribution of in-plane dipole orientations. After unloading, the system relaxes to state III$'$, where dipoles occupy four symmetry-related wells, clearly distinct from the initial state I.
These results demonstrate that the critical stress $\sigma_5^t$ associated with $\Delta \beta^t$ defines the limit for reversible piezoelectric switching. We thus define the reversible shear piezoelectric coefficient as $d_{15} = \Delta P_1^t / \sigma_5^t = 163$~pC/N.

\begin{table}[htp!]
    \centering
    \caption{Intrinsic piezoelectric strain tensor (in pC/N) calculated from MD simulations in the crystal frame, and the effective piezoelectric strain tensor ($d^{\rm eff}$) in the laboratory frame, both presented in matrix form. The largest component is highlighted in bold.}
    \begin{ruledtabular}
    \begin{tabular}{ccc}
    $d_{ij}$ & Crystal Frame (intrinsic) & Laboratory  Frame ($d^{\rm eff}$) \\
    \hline
    $\begin{pmatrix}
        d_{11} & d_{12} & d_{13} & d_{14} & \textbf{$d_{15}$} & d_{16} \\
        d_{21} & d_{22} & d_{23} & d_{24} & d_{25} & d_{26} \\
        d_{31} & d_{32} & d_{33} & d_{34} & d_{35} & d_{36}
    \end{pmatrix}$ & 
    $\begin{pmatrix}
        77 & 54 & 27 & -1 & \textbf{163} & 40 \\
        -15 & -37 & -14 & 32 & 16 & -135 \\
        25 & 11 & 14 & 0 & 45 & 0
    \end{pmatrix} $&
    $\begin{pmatrix}
        -47 & 23 & 78 & 15.6 & -7 & 13 \\
        -30 & -37 & 0 & -89 & -5 & -105 \\
        -104 & 11 & \textbf{211} & 28 & -22 & 22
    \end{pmatrix}$
    \end{tabular}
    \end{ruledtabular}
    \label{tab:dij}
\end{table}

Table~\ref{tab:dij} presents the intrinsic piezoelectric tensor, calculated in the single-crystal, single-domain limit at 300~K using the method described above. Interestingly, the longitudinal component $d_{33}$ is relatively low at 14~pC/N, while the components $d_{15} = 163$~pC/N and $d_{11} = 77$~pC/N are significantly larger in magnitude. This indicates that stress applied along the crystallographic $c$-axis induces minimal polarization change, reflecting the structural rigidity of the CdCl$_3^-$ inorganic chains.

Experimentally, as-grown ferroelectric monoclinic TMCM-CdCl$_3$ thin films typically adopt the (110) orientation [or the equivalent (100) orientation in the paraelectric phase]~\cite{Song23p2211584}. As a result, the stress applied along the film normal direction, denoted $\sigma^Z$ (aligned with the laboratory $Z$-axis), does not coincide with the crystallographic $c$-axis. Consequently, the effective piezoelectric coefficient measured along $Z$, $d^{\rm eff}_{33}$, reflects a combination of multiple components of the intrinsic piezoelectric tensor.
To quantitatively connect the intrinsic response to the experimental geometry, we perform a tensor rotation to identify the relative orientation between the stress along the film normal and polarization direction that maximizes $d^{\rm eff}_{33}$. We find that $d^{\rm eff}_{33}$ reaches a value of 211~pC/N when $\sigma^Z$ is nearly orthogonal to $[0\bar{1}2]$ or $[0\bar{1}\bar{2}]$ polarization vectors (see schematic in Fig.~S7), in excellent agreement with the experimentally measured value of 220~pC/N.  

The origin of the enhanced effective response, exceeding that of any single intrinsic component, can be understood as follows.  
Due to the relative freedom of molecular cations in the host--guest framework of TMCM-CdCl$_3$, The polarization vector does not align with any of the unit cell's crystallographic axes in the ferroelectric monoclinic phase. This anisotropy allows $\sigma^Z$ to project onto multiple stress components in the crystal frame: $\sigma^Z = 0.5\sigma_1 + 0.4\sigma_3 + 0.5\sigma_5$. As a result, the effective response includes weighted contributions from several intrinsic tensor elements, specifically $d_{11}$, $d_{13}$, $d_{15}$, $d_{31}$, $d_{33}$, and $d_{35}$, which collectively yield the giant piezoelectric response of 211~pC/N. We find that $d_{15}$ and $d_{11}$ account for about 60\% and 20\% of $d_{33}^{\rm eff}$, respectively.

The excellent agreement between theory and experiment not only validates our finite-temperature MD-based approach for computing the intrinsic piezoelectric tensor, but also provides microscopic insights into the origin of the giant response. Microscopically, the giant piezoelectric response of TMCM-CdCl$_3$ primarily arises from stress-driven, symmetry-governed stochastic hopping of molecular dipoles by 120$^\circ$. Crucially, only a small subset of local dipoles are activated, ensuring a reversible mechanical response. This gives rise to a macroscopically small-angle rotation of the net polarization vector. As shown in Fig.~\ref{fig:fig3}e, $\Delta \beta^t$ corresponds to a maximum rotation of global polarization by approximately $14^\circ$, significantly smaller than previously proposed values ($56^\circ$)~\cite{Ghosh20p207601}, setting this mechanism apart from the conventional polarization rotation mechanism~\cite{Fu00p281}. Importantly, this behavior occurs without any phase transition, in contrast to earlier interpretations~\cite{You17p306,Ghosh20p207601}.

\begin{figure*}
\centering
\includegraphics[width=1\textwidth]{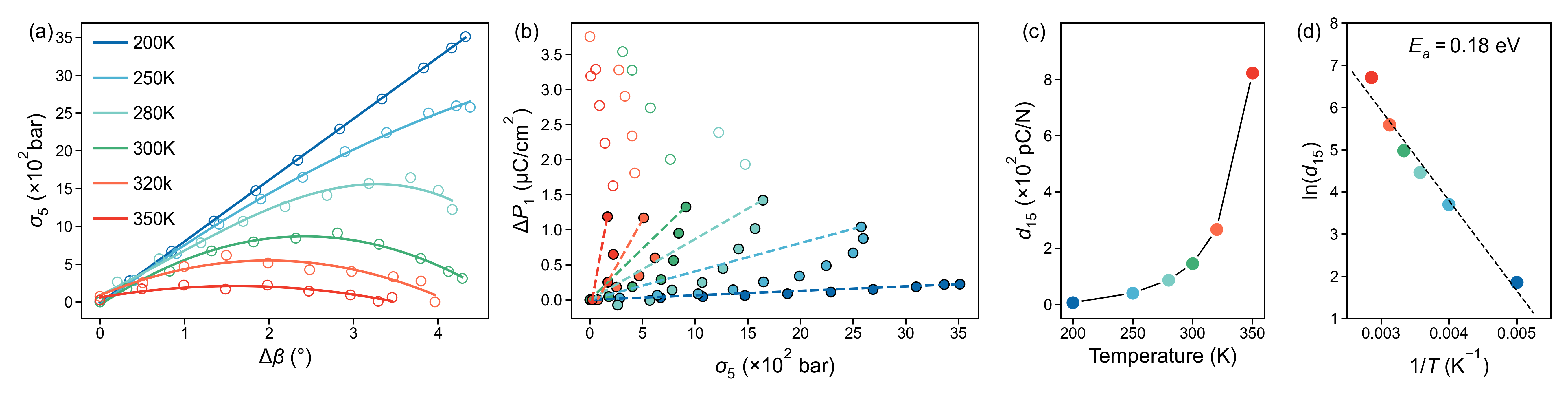}
\caption{
\textbf{Temperature-dependent piezoelectric response of TMCM-CdCl$_3$}. 
(a) Shear stress–strain curves obtained from MD simulations at temperatures ranging from 200 K to 350 K. Solid lines represent polynomial fits to the simulation data. 
(b) Corresponding polarization change $\Delta P_1$ as a function of applied shear stress $\sigma_5$ at each temperature. The piezoelectric coefficient $d_{15}$ is extracted from the reversible response region, indicated by solid markers. 
(c) Temperature dependence of $d_{15}$.
(d) Arrhenius plot of $\ln d_{15}$ versus $1/T$, with the dashed line indicating a linear fit. The slope yields an activation energy of $E_a = 0.18$~eV.
}
\label{fig:fig4}
\end{figure*}

Since the piezoelectric response of TMCM-CdCl$_3$ is governed by the stochastic hopping of molecular dipoles, a process inherently influenced by thermal fluctuations, we expect the piezoelectric coefficients to exhibit strong temperature dependence.
To investigate this, we construct $\sigma_5$--$\Delta \beta$ curves at various temperatures ranging from 200~K to 350~K (Fig.~\ref{fig:fig4}a). From these, we derive the corresponding $\Delta P_1$--$\sigma_5$ curves (Fig.~\ref{fig:fig4}b). At each temperature, the shear piezoelectric coefficient $d_{15}$ is estimated using the relation $d_{15} = \Delta P_1^t / \sigma_5^t$. As expected, $d_{15}$ shows a pronounced temperature dependence, increasing from 6~pC/N at 200~K to a giant value of 823~pC/N at 350~K.
We find that this temperature dependence follows an Arrhenius-type relationship: plotting $\ln d_{15}$ against $1/T$ yields a clear linear trend. The slope of this line corresponds to an activation energy ($E_a$) of 0.18~eV, which agrees well with the 0.21~eV rotational barrier for TMCM$^+$ cations obtained from DFT calculations (see Fig.~S3).
The Arrhenius-type temperature dependence of the piezoelectric response demonstrates that the giant effect in TMCM-CdCl$_3$ at room temperatures arises from a dynamic process governed by thermally activated hopping of molecular dipoles over an energy barrier.

As discussed above, a large applied stress can induce an irreversible structural transformation when the shear distortion exceeds $\Delta \beta^t$.  
The turnover in the stress--strain curve beyond $\Delta \beta^t$ reflects a stress-relaxation process (Fig.~\ref{fig:fig3}a), conceptually analogous to plastic deformation in metals. MD simulations reveal that this relaxation occurs through the hopping of molecular dipoles, leading to the formation of ``orientational defects," which play a similar role to dislocation defects in metals that enable plasticity~\cite{Hansen14p1681}.  
This analogy motivates further exploration of the mechanical response of TMCM-CdCl$_3$ under large shear stress or strain, which could define the material’s operational limit.

\begin{figure*}
\centering
\includegraphics[width=1\textwidth]{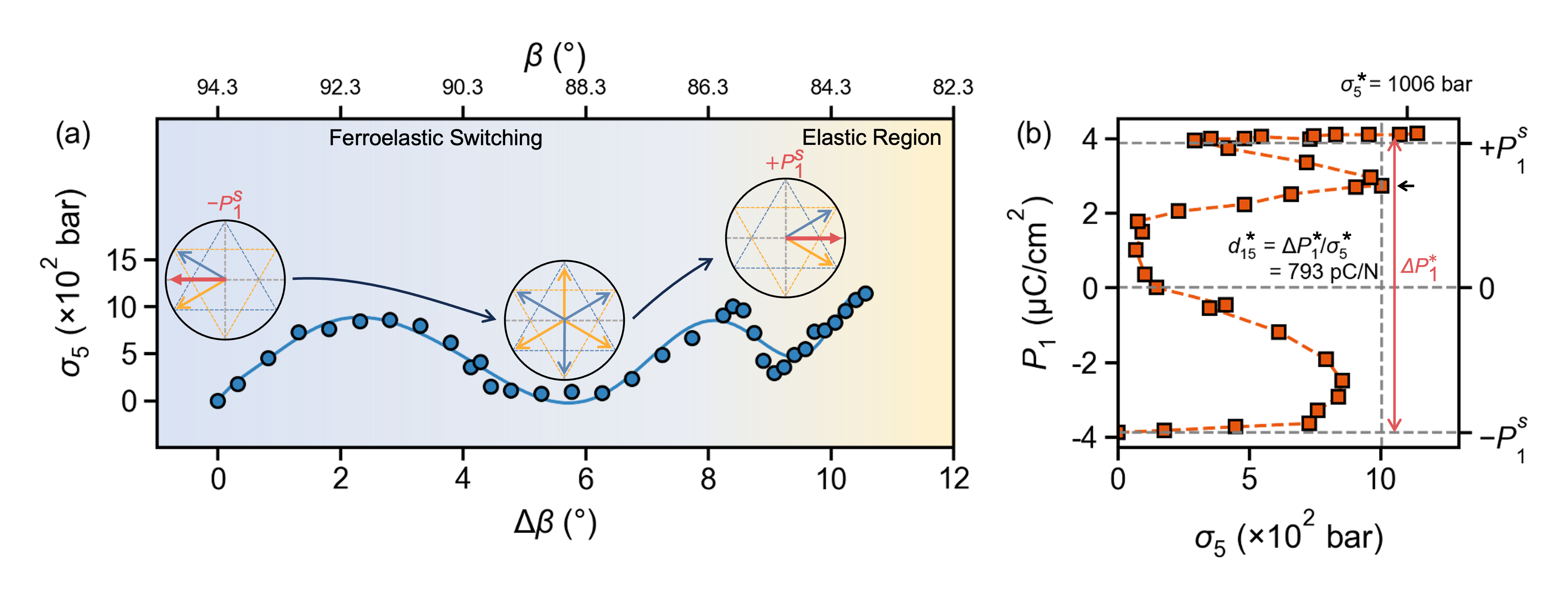}
\caption{
\textbf{Ferroelastic switching driven by large shear stress.}  
(a) Shear stress–strain curve corresponding to a full reversal of in-plane polarization ($-P_1^s \rightarrow +P_1^s$) driven by shear stress $\sigma_5$. Insets illustrate the in-plane orientations of molecular dipoles during the switching process. The stress exhibits a characteristic increase followed by a decrease, reminiscent of the plastic deformation behavior observed in metals. An intermediate state near $\beta \approx 90^\circ$ shows a nearly isotropic distribution of in-plane dipoles. Upon completing the switching, the system enters an elastic regime where stress increases linearly with strain.  
(b) Hysteresis loop showing the relationship between shear stress $\sigma_5$ and polarization $P_1$ during the switching process in (a). The maximum stress $\sigma_5^*$ defines the threshold required to complete the ferroelastic switching. The corresponding maximum (irreversible) piezoelectric response is estimated as $d_{15}^* = \Delta P_1^* / \sigma_5^*$.
}
\label{fig:fig5}
\end{figure*}

Figure~\ref{fig:fig5}a presents the full stress-strain curve, corresponding to an in-plane 180$\degree$ ferroelastic switching event, as the monoclinic angle $\beta$ is swept from 94.3$\degree$ to 85.7$\degree$.
During this transition, the stress passes through a local minimum at $\beta \approx 90 \degree$, which corresponds to an intermediate state characterized by a nearly in-plane isotropic distribution of molecular dipole orientations.
Once the switching is complete, the molecular dipoles have exhausted all available hopping pathways. Consequently, further applied strain can only be accommodated by elastic lattice distortion, causing the stress to rise sharply and linearly, a regime we refer to as the ``elastic region."
In the corresponding $P_1$--$\sigma_5$ curve shown in Fig.~\ref{fig:fig5}b, we can define another critical stress, $\sigma_5^*$, which is the \textit{minimum} stress required to complete the ferroelelastic switching process. Assuming a quasi-equilibrium process driven by $\sigma_5^*$, the system eventually equilibrates in the elastic region when the internal stress balances $\sigma_5^*$. This allows us to define an effective piezoelectric coefficient associated with ferroelastic switching: $d_{15}^* = \Delta P_1^* / \sigma_5^* = 793$~pC/N.

However, we note that this ultrahigh response is a one-time effect under a unipolar stress condition. After unloading, the system relaxes to the $+P_1^s$ state (Fig.~\ref{fig:fig5} inset). Subsequent stress application in the same direction  will only operate within the elastic regime, yielding a negligible piezoelectric response. This irreversible ferroelastic switching thus acts as a fatigue mechanism, explaining how a sample could be rendered piezoelectrically inactive if driven beyond its reversible, intrinsic limit. That said, if ferroelastic switching could be harnessed in a reversible fashion, for example, through domain wall engineering, then the piezoelectric response of TMCM-CdCl$_3$ could potentially be enhanced by over 350\%.

In summary, using finite-temperature, large-scale molecular dynamics simulations enabled by first-principles-based machine-learning force field, we
have developed a robust approach to compute the full piezoelectric tensor for organic-inorganic hybrid perovskites represented by TMCM-CdCl$_3$. This method captures key anharmonic effects associated with the thermally driven order–disorder behavior of molecular dipoles.
We reproduce experimental giant piezoelectric coefficient, demonstrating that the effect is dominated by intrinsic components like $d_{15}$ and $d_{11}$, which arise from stochastic in-plane hopping of molecular dipoles. 
This hopping is not random: the local C$_3$-symmetric halogen-bonding network between TMCM$^+$ cations and the inorganic host framework creates a rotational potential energy landscape with three symmetry-equivalent minima. 
The Arrhenius-type temperature dependence of the piezoelectric coefficients further corroborates the critical role of dipole hopping. The mechanism of discrete, localized hopping is fundamentally different from the continuous, bulk polarization rotation that drives the response in conventional piezoceramics like PZT near a morphotropic phase boundary~\cite{Guo2000p5423,Fu00p281}. Importantly, the reversibility of the piezoelectric response in TMCM-CdCl$_3$ also relies on the activation of only a small subset of molecular dipoles. We also identify in-plane 180$^\circ$ ferroelastic switching as both a potential fatigue mechanism and a promising route to further enhance the electromechanical response if rendered reversible. Our work underscores the potential of engineering host–guest halogen bonds to precisely tailor the rotational energy landscape of molecular dipoles for improved performance.
\\

\begin{acknowledgments}
We acknowledge the supports from National Natural Science Foundation of China (92370104).
The computational resource is provided by Westlake HPC Center. 
\end{acknowledgments}

\bibliography{SL.bib}

\clearpage
\newpage

\clearpage
\newpage

\clearpage
\newpage

\clearpage
\newpage

\end{document}